# FAST OR ACCURATE? GOVERNING CONFLICTING GOALS IN HIGHLY AUTONOMOUS VEHICLES

A. FEDER COOPER† AND KAREN LEVY†

*The tremendous excitement around the deployment of autonomous vehicles (AVs) comes from their purported promise. In addition to decreasing accidents, AVs are projected to usher in a new era of equity in human autonomy by providing affordable, accessible, and widespread mobility for disabled, elderly, and low-income populations. However, to realize this promise, it is necessary to ensure that AVs are safe for deployment, and to contend with the risks AV technology poses, which threaten to eclipse its benefits. In this Article, we focus on an aspect of AV engineering currently unexamined in the legal literature, but with critical implications for safety, accountability, liability, and power. Specifically, we explain how understanding the fundamental engineering trade-off between accuracy and speed in AVs is critical for policymakers to regulate the uncertainty and risk inherent in AV systems. We discuss how understanding the trade-off will help create tools that will enable policymakers to assess how the trade-off is being implemented. Such tools will facilitate opportunities for developing concrete, ex ante AV safety standards and conclusive mechanisms for ex post determination of accountability after accidents occur. This will shift the balance of power from manufacturers to the public by facilitating effective regulation, reducing barriers to tort recovery, and ensuring that public values like safety and accountability are appropriately balanced.*

† PhD Candidate, Department of Computer Science, Cornell University.

† Assistant Professor, Department of Information Science, Cornell University; Associate Faculty, Cornell Law School. We thank the following people for discussion and feedback on prior versions of this paper: Jack Balkin, Solon Barocas, Ken Birman, Fernando Delgado, Chris De Sa, James Grimmelmann, Steve Hilgartner, Jon Kleinberg, Helen Nissenbaum, Fred B. Schneider, and Ed Walters. We also thank the Artificial Intelligence, Policy, and Practice Initiative at Cornell University and the (Im)perfect Enforcement Conference at Yale Law School's Information Society Project for workshopping earlier versions of this work, and the John D. and Catherine T. MacArthur Foundation and the Digital Life Initiative at Cornell Tech for their support.

## INTRODUCTION

Perhaps no technology has aroused greater excitement in recent years than the development and commercialization of autonomous vehicles (AVs). In addition to potentially improving safety,[1] AVs have the potential to support equity in human mobility

1. The National Highway Traffic Safety Administration (NHTSA) estimates that "human factors," such as speeding, intoxication, and inattention, contribute to an astounding 94% of traffic accidents. *See* NATIONAL HIGHWAY TRAFFIC SAFETY ADMINISTRATION, U.S. DEP'T OF TRANSP., DOT HS 812 115, CRITICAL REASONS FOR CRASHES INVESTIGATED IN THE NATIONAL MOTOR VEHICLE CRASH CAUSATION SURVEY 1 (Feb. 2015), https://crashstats.nhtsa.dot.gov/api/public/viewpublication/812115 [https://perma.cc/Q5UB-M87Z]. With approximately 35,000 accident-related deaths per year, if AVs manage to prevent the majority of human-factors related accidents, on the and autonomy by providing unprecedented affordable, accessible, and widespread mobility for disabled, elderly, and low-income populations.[2] As of 2016, nineteen major car manufacturers have announced plans to develop AV technology in coming years.[3] By 2040, 66% of cars are expected to feature at least some autonomous driving capabilities.[4]

However, to realize this promise, it is necessary to ensure that AVs are safe for deployment—and to contend with the risks AVs pose,[5] which threaten to eclipse their potential benefits. Notably, many scholars have drawn attention to the concern that the machine learning algorithms used in AVs make decisions in ways that are not easily explainable to human regulators—that it will be impossible to assess *why* an AV made a particular mistake, thus muddling the ability to determine accountability[6] and raising novel order of tens of thousands of lives could be saved in the US each year. *See* NATIONAL HIGHWAY TRAFFIC SAFETY ADMINISTRATION, U.S. DEP'T OF TRANSP., DOT HS 812 376, SUMMARY OF MOTOR VEHICLE CRASHES 2 (Feb. 2020), https://crashstats.nhtsa.dot.gov/Api/Public/ViewPublication/812376 [https://perma.cc/9BF4-EKEE]. Worldwide, some estimates for 2035–2045, the decade in which AV technology is projected to reach widespread deployment, suggest that 585,000 lives will be saved. ROGER LANCTOT, ACCELERATING THE FUTURE: THE ECONOMIC IMPACT OF THE EMERGING PASSENGER ECONOMY 6 (Strategy Analytics, 2017), https://newsroom.intel.com/newsroom/wp-content/uploads/sites/11/2017/05/passenger-economy.pdf [https://perma.cc/2Z85-JTGH].

2. *See* LANCTOT, *supra* note 1, at 15. For a more comprehensive analysis of the potential benefits of large-scale deployment of AV technology, *see generally* SCOTT SMITH ET AL., BENEFITS ESTIMATION FRAMEWORK FOR AUTOMATED VEHICLE OPERATIONS, REPORT PREPARED FOR U.S. DEP'T OF TRANSP., INTELLIGENT TRANSPORTATION SYSTEMS JOINT PROGRAM OFFICE (2015), https://rosap.ntl.bts.gov/view/dot/4298 [https://perma.cc/F4NJ-3D2Q].

3. AMERICAN ASSOCIATION FOR JUSTICE, DRIVEN TO SAFETY: ROBOT CARS AND THE FUTURE OF LIABILITY 17 (2017), https://www.justice.org/resources/research/driven-to-safety-robot-cars [https://perma.cc/HEA7-463Y].

4. By 2040, McKinsey projects 66% of kilometers driven will be by AVs. It takes approximately 15 years for cars in use to completely turn over, so there will be a mix of automation technologies (at different levels) for many years into the future. MCKINSEY CENTER FOR FUTURE MOBILITY, THE FUTURE OF MOBILITY IS AT OUR DOORSTEP: COMPENDIUM 2019/2020 47–48 (2020), https://www.mckinsey.com/~/media/McKinsey/Industries/Automotive and Assembly/Our Insights/The future of mobility is at our doorstep/The-future-of-mobility-is-at-our-doorstep.ashx [https://perma.cc/T9AB-KKA3]. For example, while electronic differential locking systems have existed since the 1990s, it will take until 2032 for 95% of actively-used cars to possess the feature. *See* AMERICAN ASSOCIATION FOR JUSTICE, *supra* note 3, at 8–9.

5. *E.g.*, Neal E. Boudette, *Tesla Says Autopilot Makes Its Cars Safer. Crash Victims Say It Kills*, N.Y. TIMES (July 5, 2021), https://www.nytimes.com/2021/07/05/business/tesla-autopilot-lawsuits-safety.html [https://perma.cc/3UNE-EMU3] (describing the recent lawsuit against Tesla concerning safety risks in its autonomous driving technology).

6. *See* Joshua A. Kroll et al., *Accountable Algorithms*, 165 U. PA. L. REV. 633, 639 (2017) (showing a legally-focused primer concerning the explainability and accountability of automated decision algorithms in general). For accountability in the context of autonomous vehicles, *see generally* Sven Nyholm and Jilles Smids, *The Ethics* legal questions on a range of issues, from the Fourth Amendment automobile exception[7] to tort liability[8] and the regulation of the automotive insurance industry.[9]

In this Article, we focus on a deeper and underexplored aspect of AV engineering—one unexamined in the legal literature, but with critical implications for safety, accountability, liability, and power. Specifically, we explain how understanding the fundamental engineering trade-off between the accuracy and speed of decision-making is critical for policymakers to regulate the uncertainty and risk inherent in AV systems.

Autonomous vehicles are *distributed systems*: networks of sensors—for example, GPS, cameras, LIDAR, and radar—that record different data and work together to inform the car's behavior.[10] Among them, these devices may have an inaccurate view of that environment: for example, one camera may detect a pedestrian up ahead, while another camera may not yet have the pedestrian in its view. These different views need to be coordinated in order to build a coherent view of the AV's surroundings. This coordination takes time to compute, but decisions need to be made very quickly in order to be useful. Engineers must decide how accurate is accurate *enough* for the AV to make a decision—recognizing that waiting for too much certainty (by allowing for the reconciliation of inaccuracies) can itself create risk. In other words, AVs exhibit a trade-off inherent to distributed systems[11] (and many

---

*of Accident-Algorithms for Self-Driving Cars: an Applied Trolley Problem?, 19* ETHICAL THEORY & MORAL PRACTICE 1275, (2016); *see also* Madeleine Clare Elish, *Moral Crumple Zones: Cautionary Tales in Human-Robot Interaction*, 4 ENGAGING SCI., TECH., & SOC'Y 40 (2019). Harry Surden and Mary-Anne Williams, *Technological Opacity, Predictability, and Self-Driving Cars*, 38 CARDOZO L. REV. 121, 130 (2016) ("Autonomous vehicles occupy a middle ground that has little or no comparator today among moving entities. On one hand, their automated movements are not limited to highly circumscribed, repetitive routes, as are elevators. Rather, autonomous vehicles are capable of driving on ordinary roads, going nearly anywhere a human driver might go. On the other hand, their movement choices are made by computer systems, not by humans. Their movements are, therefore, not intuitively revealed through cognitive introspection and projection.").

7. *See generally* Lindsey Barrett, *Herbie Fully Downloaded: Data-Driven Vehicles and the Automobile Exception*, 106 GEORGETOWN L.J. 181 (2017).

8. AMERICAN ASSOCIATION FOR JUSTICE, *supra* note 3, at 7.

9. *See generally* Kenneth S. Abraham & Robert L. Rabin, *Automated Vehicles and Manufacturer Responsibility for Accidents: A New Legal Regime for a New Era*, 105 VIRGINIA L. REV. 127 (2019).

10. A. Feder Cooper et al., *Accuracy-Efficiency Trade-Offs and Accountability in Distributed ML Systems*, EQUITY AND ACCESS IN ALGORITHMS, MECHANISMS, AND OPTIMIZATION (EAAMO '21) Article 4 (2021), at 6.

11. This trade-off is in fact relevant across computing, including machine learning (ML). For a treatment of this subject in relation to distributed ML systems, *see id*.

other policy contexts): the trade-off between *how fast* decisions are made and *how accurate* those decisions can be.

While this trade-off may seem like a technical implementation detail, such engineering trade-offs entail broader impacts; they are implicated in tensions among social values like safety, efficiency, and equity, which policymakers must navigate and balance in regulating these systems.[12] Therefore, it is crucial to make this technical trade-off legible to policymakers seeking to regulate AV systems, as this legibility is essential for successfully navigating and balancing tensions among social values.

In this Article, we examine how we can create tools that enable policymakers to assess how the trade-off is being implemented, thereby providing an actionable path for informing the regulation of AV implementation decisions. Such tools will facilitate opportunities for developing concrete *ex ante* AV safety standards and conclusive mechanisms for *ex post* determination of accountability after accidents occur. Importantly, these *ex post* mechanisms can help to diagnose the nature of accidents, and will help distinguish uncertainty due to the trade-off from other sources of uncertainty that pose potential safety risks, such as software bugs and hardware malfunctions. The National Highway Transportation Safety Administration (NHTSA, the federal automotive regulatory authority)[13] currently relies on automotive manufacturers to perform *ex ante* self-certification, while taking on the burden and cost of properly collecting data and analyzing *ex post* if a recall is warranted.[14] NHTSA therefore needs effective tools to exercise its *ex post* recall authority for AVs, as a balance on the power that car manufacturers have to potentially self-certify AV technology that is in fact not safe to deploy. In other words, NHTSA needs tools to contend with the unbalanced distribution of power between AV manufacturers and individual AV users—an

12. *See* Jake Goldenfein et al., *Through the Handoff Lens: Competing Visions of Autonomous Futures*, 35 BERKELEY TECH. L.J. 835, 838 (2020); National Highway Traffic Safety Administration, *Federal Automated Vehicles Policy: Accelerating the Next Revolution in Roadway Safety*, September 2016, at 26–27 (hereinafter FAVP).

13. The federal government is responsible for regulating motor vehicles and equipment, while states tend to be responsible for regulating the human driver and other operations. In the case of AVs, these two categories lose their traditional distinction, and overlap in important ways. *See* FAVP, *supra* note 12, at 17–18.

14. On self-certification, *see* FAVP, *supra* note 12, at 71–72. On the dearth of *ex ante* regulatory authority at NHTSA, *see supra* note 1, at 11 ("[T]here is currently no specific federal legal barrier to an HAV being offered for sale."). Policy set during the Trump administration only suggests that AV manufacturers submit "Voluntary Safety Self-Assessments" concerning meeting Federal Motor Vehicle Safety Standards (FMVSS). These assessments are neither required nor does NHTSA have any mechanism to compel them. *See* LEE VINSEL, MOVING VIOLATIONS: AUTOMOBILES, EXPERTS, AND REGULATIONS IN THE UNITED STATES 296–97 (2019).

imbalance that may particularly harm the marginalized populations AVs promise to benefit.

Providing an effective *ex post* counterbalance to manufacturer self-certification has historically proven challenging and will become even more so with AVs. In the past, automotive manufacturers have taken advantage of the lack of *ex ante* oversight to knowingly deploy faulty technology, including defective ignition switches and airbags.[15] Such offenses have become more common and sophisticated as car technology has evolved to be increasingly computerized, even prior to the advent of autonomous features; computerization has facilitated more nuanced evasion of standards.[16] Policymakers need tools to reason precisely about the inherent trade-off between accuracy and speed so that they can more effectively prevent car manufacturers from concealing misconduct. This would also increase public accountability over the engineering decisions made in AV systems and ensure that public values are represented in their design so that AV technology is sufficiently safe to deploy.

We proceed as follows. In Part I, we provide technical background on the trade-off between *accuracy*[17] and *speed* in distributed computing systems. We illustrate the intuition for this trade-off, define the relevant technical terms in relation to familiar user experiences with the Internet,[18] and then apply these definitions to explain how AV technology presents unique challenges in terms of implementing the trade-off. In Part II we clarify that trade-off implementation choices present a valid site for regulatory intervention, as different choices implicate balancing tensions between broader social values, like safety and efficiency. Policymakers already successfully navigate comparable tensions in other high-stakes regulatory domains, including public health; we therefore argue that an AV's resolution of the trade-off could similarly be subject to regulatory scrutiny. We offer a path forward for concrete regulatory interventions for AVs in Part III, arguing that, with the right tools in place, it is possible to enable more effective democratic governance of complex systems like AVs. More

15. *See* AMERICAN ASSOCIATION FOR JUSTICE, *supra* note 3, at 17–21, 24, 33–34. For example, GM concealed an ignition switch defect for 10 years, which led to the deaths of at least 124 people. AMERICAN ASSOCIATION FOR JUSTICE, *supra* note 3, at 19.

16. *See* VINSEL, *supra* note 14, at 270–71.

17. More formally, this is called *consistency* in distributed systems. We use the term accuracy because it captures the necessary meaning for the purposes of this Article and is more familiar. However, where appropriate in technically-focused footnotes, we will use the formal term.

18. *See generally* Barry M. Leiner et al., *A Brief History of the Internet,* 39 SIGCOMM COMPUT. COMMUN. REV. 22 (2009) (explaining the early history of the internet, which was the first large-scale distributed computer system.).

specifically, we frame our discussion in terms of how such tools could benefit the development of both *ex ante* AV safety standards, which have the potential to help prevent accidents, and *ex post* interventions concerning recording data for audits, attribution of errors, and tort liability when bad outcomes invariably occur.

## I. THE INHERENT TRADE-OFF BETWEEN ACCURACY AND SPEED

Before addressing the policy implications of the accuracy-speed trade-off for AVs, it is important to provide clear definitions of the technical concepts that underlie it. In this section, we define the trade-off precisely and illuminate its technical underpinnings via various computer systems examples. We explain the three key concepts—*distributed system*, *accuracy*, and *speed*—on which the rest of our discussion relies. We begin by providing an intuition of these concepts via an extended metaphor, then define the trade-off between accuracy and speed in relation to familiar user experiences on the Internet, and lastly describe the particular complexities of the trade-off for the implementation of AV systems.

### *A. Beginning with an Intuition*

A *distributed system* consists of several computers, also called nodes, that are spatially separated and communicate with each other.[19] The computers can work together to solve problems: each computer has its own data and performs its own computations, and, when necessary, it shares those data and computation results with other computers in the network. If a computer needs data from another computer in order to execute a computation, it can request the data from that computer.

In such systems, where the different nodes separately ingest and process different data, it is nontrivial for all the nodes to agree at a particular moment in time about the state of the overall system; the nodes can have inaccurate views about the overarching system's current state. For the overall system to make useful decisions, it is often important to reconcile these inaccuracies (at least to a certain extent). This difficult task of forming a coherent, holistic understanding of a dynamic environment relates to a classic

19. For a more formal definition of a distributed system, *see* Leslie Lamport, *Time, Clocks, and the Ordering of Events in a Distributed System*, 21 COMM. ACM 558, 562–63 (1978). We use the term *distributed systems* to include networks of connected computers that integrate data from more than one distinct computing device. Practically speaking, a single computer such as a laptop could also be broadly conceived as a distributed system; such a device has the ability to run multiple processes, each of which is an instance of a running program. *See generally* REMZI H. ARPACI-DUSSEAU & ANDREA C. ARPACI-DUSSEAU, OPERATING SYSTEMS: THREE EASY PIECES 25–27 (2018).

problem in distributed systems. Distributed systems researchers have long recognized an inherent trade-off between *accuracy* and *speed*—between waiting to make a completely informed decision and making a decision fast enough for it to be useful. A canonical paper in the field describes the intuition behind this complex problem by way of analogy.[20] It illustrates the difficult task of integrating an accurate view in a distributed system by comparing it to several photographers trying to capture a single large image of a sky full of birds:

> [A] group of photographers [is] observing a panoramic, dynamic scene, such as a sky filled with migrating birds—a scene so vast that it cannot be captured by a single photograph. The photographers must take several snapshots and piece the snapshots together to form a picture of the overall scene. The snapshots cannot all be taken at precisely the same instant because of synchronization problems. Furthermore, the photographers should not disturb the process that is being photographed; for instance, they cannot get all the birds in the heavens to remain motionless while the photographs are taken. Yet, the composite picture should be meaningful. The problem before us is to define *meaningful* and then to determine how the photographs should be taken.[21]

This meaningful picture should be accurate. We need to account for each bird exactly once—we do not want to undercount or overcount them when the photos are stitched together. We can control how much time we spend on this process of stitching together the images. If performed slowly and methodically, it should be possible to account correctly for each bird, producing an image consistent with the actual sky; if completed hastily, inaccuracies such as duplicated birds could appear in the resulting image. In other words, there is an inevitable trade-off between capturing a perfect, accurate image and how much time is spent producing that image. Depending on the time constraints, it is not always possible to produce a perfectly accurate image.

Systems that aim to provide real-time responsiveness—that is, that aim to minimize how much time is spent executing computations—will necessarily require a sacrifice in the degree of accuracy. This poses a crucial challenge for a system like an

20. *See* K. Mani Chandy & Leslie Lamport, *Distributed Snapshots: Determining Global States of Distributed Systems*, 3 ACM TRANSACTIONS ON COMPUT. SYS. 63, 69–71 (1985).

21. *Id.* at 64.

autonomous car, which ideally wants to guarantee both speed and accuracy. For safety, the car needs to build snapshots of its environment in real time, but also must prioritize accuracy when constructing them. The trade-off between speed and accuracy therefore presents a significant and underappreciated problem for technologies like AVs—and, importantly, for the policies we build around distributed systems. We cannot build distributed computing systems that simultaneously work as *quickly as possible* and also reflect the world as *accurately as possible*. Instead, system designers make technical choices that necessarily prioritize between these competing goals.

### *B. The Trade-Off at Play in Familiar Settings*

The *accuracy-speed trade-off*,[22] and the different design choices made to deal with it, is not relevant only for emerging technology like AVs. Rather, different implementations of the trade-off are present everywhere in modern technological systems—and affect our everyday experiences on the Internet in ways we may not realize.

In fact, this tension is one of the most important issues designers must consider when designing systems. When a request is made for data in a distributed system, multiple nodes can be contacted to get a picture of the correct current system state. If all of the nodes are contacted, then it is possible to reconcile their views to determine the state. However, contacting all of the other nodes and figuring out an accurate picture takes time; it is a high-latency (slow) interaction. If fewer nodes are contacted, it takes less time but there is a higher probability that the responses will fail to create an accurate picture of the system's state. In other words, there is a *spectrum* in the trade-off between accuracy and speed; as shown in Figure 1, it is not all-or-nothing.[23] As a result, there is an opportunity for flexibility—for an application to implement the trade-off in a manner that is appropriate to its respective goals and priorities.

---

22. CAP (Consistency, Availability, Partition Tolerance) is a related, though more contentious, concept in distributed computing. *See generally* Eric Brewer, *CAP Twelve Years Later: How the "Rules" Have Changed*, 45 COMPUT. 23, 24 (2012) (describing the relationship between these concepts); Daniel Abadi, *Consistency Tradeoffs in Modern Distributed Database System Design: CAP Is Only Part of the Story*, 45 COMPUT. 37, 38–40 (2012) (also discussing this relationship). *See* DAVE CLARK, DESIGNING AN INTERNET 227–29 (2018) (showing a simple characterization of the relationship between availability (the A in CAP, in the citations above) and latency).

23. More strongly consistent protocols involve contacting multiple nodes and confirming that they agree or resolving conflicts before returning a response, which incurs a latency cost (that is, it takes time). *See generally* Werner Vogels, *Eventually Consistent*, 52 COMM. ACM 40, 42 (2009).

User interactions on social media provide an accessible example of this technical problem. Social media aims to be constantly responsive in order to maintain user engagement. To maintain the fluidity of user experience, users must perceive that the actions they take "register" immediately, without having to wait for even a short time. In other words, social media sites are time-sensitive;[24] they prioritize speed, which comes with a cost to accuracy. As a result, these systems sometimes exhibit odd behavior due to inaccuracies between computers in the system. Delays between when different nodes on the site receive updates can lead to timing issues, which manifest bizarrely in the website's user experience—for example, seeing comments out-of-order on a newsfeed, or trying to like a post only to find that the creator has deleted it. These irregularities are the result of the decision to prioritize the responsiveness of social media sites over the accuracy of the system.

In other consumer contexts, accuracy is prioritized over speed. For example, withdrawing money from an ATM is a high-fidelity process that is not instantaneous; it takes time to validate the presence of funds (i.e., to ensure that all computers in the network are aware of the correct current balance) and to update the balance throughout the system. Figure 1 illustrates some other common implementations of the trade-off in various computing systems.

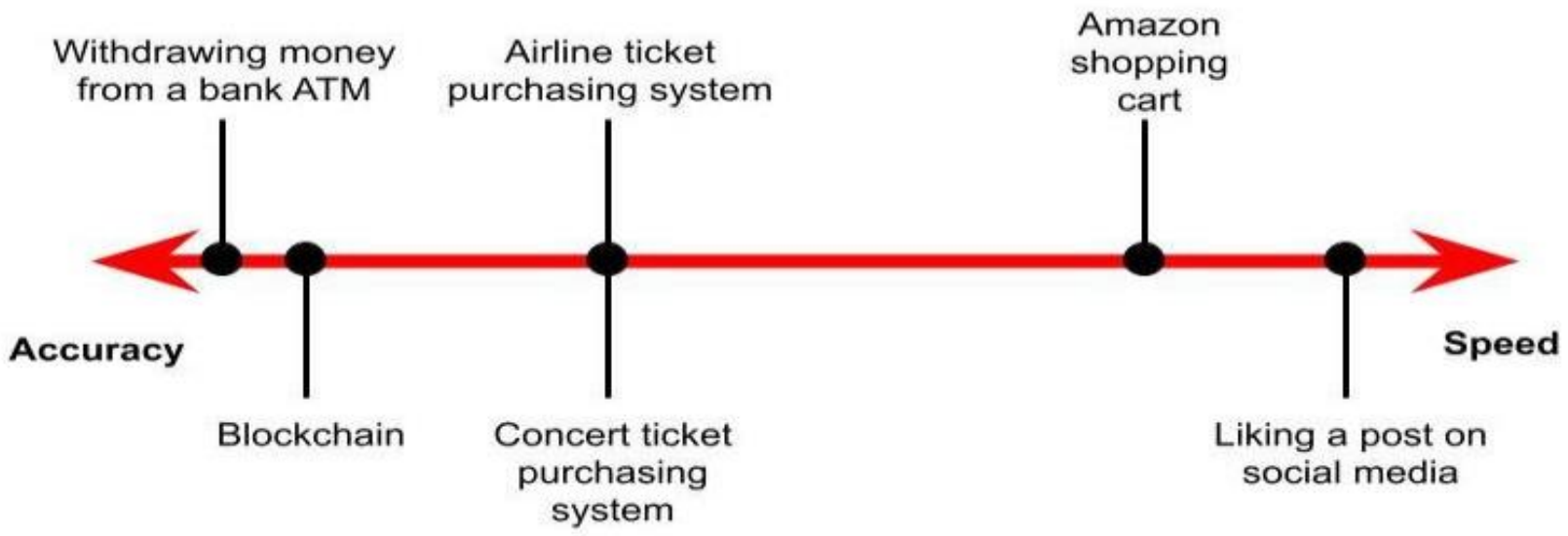

24. Speed is critical in online transactions, a 100-millisecond delay can drastically reduce the probability that a customer will return to a website. Abadi, *supra* note 21, at 38. *See also* Peter Bailis et al., *Probabilistically Bounded Staleness for Practical Partial Quorums*, 5 PROC. VLDB ENDOWMENT 776, 776 (2012) (finding that for Amazon, 100 milliseconds of extra latency resulted in a 1% drop in sales); Haonan Lu et al., *Existential Consistency: Measuring and Understanding Consistency at Facebook*, PROC. 25TH SYMP. ON OPERATING SYS. PRINCIPLES (SOSP '15) 295, 295 (2015) ("[W]eaker forms of consistency … [create] . . . user-visible *anomalies*, i.e., strange behavior that defies user expectations. A common example is out-of-order comments on a social network post, e.g., Alice comments on a post, Bob comments on the same post after seeing Alice's comments, and then Charlie sees Bob's comment appear before Alice's.").

Figure 1. A simplified visualization of the spectrum between accuracy and speed for a variety of distributed systems applications.[25]

### *C. The Trade-Off's Implications for AVs*

As is clear from these examples, this trade-off has an enormous impact on the high-level behavior of distributed systems. However, the technical choices that underlie this behavior involve low-level decisions—decisions made deep in the system's software and hardware—which determine how these systems behave and the ensuing interactions that users have with them. Autonomous vehicles will also have to contend with this trade-off; low-level implementation details concerning accuracy and speed will directly impact an AV's overarching behavior.

In this section, we highlight why the accuracy-speed trade-off is uniquely challenging for AVs in comparison to the more familiar examples described above. We focus our analysis on *highly* autonomous vehicles (HAVs).[26] "AV" is a fairly generic term that applies to vehicles with different degrees of autonomy,[27] with

25. Amazon's shopping cart favors speed over accuracy when a user adds items to their shopping cart. Browsing may show certain items as available, which can be added to the cart, but these items are not necessarily accurate with the actual, available inventory. This can become clear at checkout, where accuracy is favored over speed, attempting to complete a purchase is not instantaneous and may show that an item in the cart is in fact unavailable. This was a common experience for many consumer goods at the beginning of the COVID-19 pandemic. *See* Giuseppe DeCandia et al., *Dynamo: Amazon's Highly Available Key-Value Store*, PROC. OF THE 21ST ACM SIGOPS SYMP. ON OPERATING SYS. PRINCIPLES (SOSP '07) 205 (2007). In contrast, blockchain technology favors accuracy over speed. Blockchain is a distributed system that manages a transaction ledger. Every node in the system is responsible for keeping track of the transaction ledger and consistency between nodes (i.e., accuracy) is extremely important for maintaining correctness. This makes the system slow: In the Bitcoin system it takes up to ten minutes for a pending transaction to execute. *See Average Time to Mine a Block in Minutes*, DATA.BITCOINTY.ORG, https://data.bitcoinity.org/bitcoin/block_time/5y?f=m10&t=l [https://perma.cc/P2DN-VXDV] (last visited Feb. 26, 2022). In other words, the entire blockchain system is purposefully reliant on being slow—what Ohm and Frankle term "desirable inefficiency." *See* Paul Ohm & Jonathan Frankle, *Desirable Inefficiency*, 70 FLA. L. REV. 777, 777 (2018). *See generally* Satoshi Nakamoto, *Bitcoin: A Peer-to-Peer Electronic Cash System* (2008), http://bitcoin.org/bitcoin.pdf [https://perma.cc/R36J-PHQF].

26. NHTSA defines an AV as follows: "An automated vehicle system is a combination of hardware and software (both remote and on-board) that performs a driving function, with or without a human actively monitoring the driving environment." U.S. DEP'T OF TRANSP., FEDERAL AUTOMATED VEHICLES POLICY 10 (2016). Because of our analytical choice, we can use AV interchangeably with HAV, subsetting to the part of the definition in which the human does not actively monitor the environment.

27. There are six internationally recognized, mutually exclusive levels of automation created by the standards organization SAE International: Level 0 (no different types of partial autonomy for specific driving features in between. HAVs are AVs that do not rely on human interaction to perform any driving.[28] More technically, the human passenger does not perform any aspect of any dynamic driving task (DDT) in any operational design domain (ODD).[29]

One can think of the AV as a distributed system, in which the nodes are multiple distinct sensors—such as GPS, cameras, and LIDAR[30] detectors—that many times per second generate an enormous amount of data about road conditions, traffic signs, and the presence of pedestrians and other obstacles. Each of these sensors is, technically, a separate computational device, but they are connected together and can combine their individual data for the car to make unified decisions about its behavior—whether to apply the brakes, accelerate, or change directions. Because the sensors operate independently, they might also provide conflicting information, both in terms of *what* environmental state they detect and *when* they detect it. The car will have to resolve these conflicts to make coherent decisions; it must attempt to rapidly integrate myriad, potentially conflicting sources of information to produce an

---

automation; the human driver performs all functions) to Level 5 (full automation; AV performs all driving tasks under all conditions). *See* SAE INTERNATIONAL, SURFACE VEHICLE RECOMMENDED PRACTICE. TAXONOMY AND DEFINITIONS FOR TERMS RELATED TO DRIVING AUTOMATION SYSTEMS FOR ON-ROAD MOTOR VEHICLES, STANDARD J3016 4, 8, 24–28 (2021) (hereinafter "SAE International"); *See also* Kenneth S. Abraham & Robert L. Rabin, *Automated Vehicles and Manufacturer Responsibility for Accidents: A New Legal Regime for a New Era*, 105 VIRGINIA L. REV. 127, 128 (2019); *id.* at 149 (showing partially autonomous features common in many vehicles on the road today, including electronic stability control, automatic emergency braking, and lane keeping assistance); *id.* at 149–150 (explaining certain features are "active safety systems"; they are not features engaged in a sustained fashion, but rather only are momentary interventions).

28. SAE Levels 3-5 qualify as HAVs, but there is some debate concerning SAE 3, as it can frequently rely on the human driver for fallback. Narrowing our focus to HAVs enables us to focus on the complexities of the accuracy-speed trade-off solely in the context of the AV system, without attending to specific edge cases concerning the interplay between the trade-off and fallback to a human driver. Limiting our discussion in this way also enables us to sidestep debates around whether partially autonomous cars will ever be safe for large-scale deployment. *See* MYRA BLANCO ET. AL., U.S. DEP'T OF TRANSP., HUMAN FACTORS EVALUATION OF LEVEL 2 AND LEVEL 3 AUTOMATED DRIVING CONCEPTS 104, 136 (2015), https://www.nhtsa.gov/sites/nhtsa.gov/files/812182_humanfactorseval-l2l3-automdrivingconcepts.pdf [https://perma.cc/8PRG-CHVD] (explaining that, on average, it takes a human driver 17 seconds to fully regain control—an amount of time during which a car driving at 60 mph would travel over a quarter mile).

29. *See* SAE International, *supra* note 27, at 4–6.

30. Though AV technology is quickly developing, AVs generally have in common certain types of hardware such as LiDAR or video-capable cameras, and others among them. LIDAR is a type of remote sensing technology that is more precise than sonar. *See, e.g.*, Rui Qian et al., *End-to-End Pseudo-LiDAR for Image-Based 3D Object Detection*, 2020 IEEE/CVF CONF. ON COMPUTER VISION & PATTERN RECOGNITION 5881 (2020) [hereinafter CVPR].

accurate understanding of the environment with which it interacts. The AV will have to make choices given necessarily imperfect information, due to inconsistencies across time and across data sources, and these choices have high stakes: stopping abruptly may injure its passenger, while failing to do so may harm a pedestrian.

This problem becomes even more complex when an AV also incorporates information from other sources—including smart highway devices, like dynamic traffic lights and CCTV (Vehicle-to-Infrastructure, or V2I), and from other AVs (Vehicle-to-Vehicle, or V2V).[31] V2V communication allows multiple AVs to share data about the surrounding environment. In other words, in addition to viewing a single autonomous car as a distributed system, V2V will allow us to treat a *group* of communicating autonomous cars as a distributed system. Together, V2V and V2I promise to help AVs produce richer pictures of the driving environment through additional collecting and sharing of distributed data. However, the use of additional sensors presents the potential for more inter-device inaccuracy; an AV will not only have to resolve inaccuracies from within its own sensor system, it will also have to contend with inaccuracies due to information-sharing from other AVs and sensors, which will also require even more time to resolve. This clearly presents a difficult tension to navigate: AVs need to incorporate various sensors to make accurate decisions in their dynamic environment, *and* those decisions are also extremely time-sensitive. In other words, maintaining *both* accuracy and speed is important, because both implicate overall AV safety. Given the inherent trade-off between the two, it is not immediately clear how an AV can sacrifice either without compromising overall system safety.

The dynamic nature of the driving environment demonstrates why the implications of the trade-off are quite different for AVs than for applications like social media or buying a concert ticket. AVs have a very high degree of (fairly unrestricted) mobility, and therefore must perform in changing, variable environments.[32] In the standard technical language:

31. It is generally accepted that V2V and V2I are necessary for HAVs. Federal Motor Vehicle Safety Standards: V2V Communications, 82 Fed. Reg. 3854 (proposed Jan. 12, 2017); *U.S. Department of Transportation Issues Advance Notice of Proposed Rulemaking to Begin Implementation of Vehicle-to-Vehicle Communications Technology*, U.S. DEP'T OF TRANSP. (Aug. 18, 2014), https://www.transportation.gov/briefing-room/us-department-transportation-issues-advance-notice-proposed-rulemaking-begin [https://perma.cc/62W2-9P99]. For definitions of V2V and V2I, *see* FAVP, *supra* note 12, at 5; *see generally* 49 C.F.R. § 571 (2022).

32. *See* Surden & Williams, *supra* note 6, at 130 (contrasting highly mobile AVs with circumscribed mobile systems, like elevators).

> the ODD, which may vary for each HAV system, will define the conditions in which [a] function is intended to operate with respect to roadway types, geographical location, speed range, lighting conditions for operation (day and/or night), weather conditions, and other operational domain constraints.[33]

The various conditions and constraints of different ODDs may require different implementations of the accuracy-speed trade-off in order for the AV to perform its driving functions adequately. In other words, unlike social media platforms or ATMs, AVs present a unique challenge, because the appropriate trade-off implementation *might change depending on the environment.* The safest trade-off between accuracy and speed for an AV could vary based on context. And, of course, in the realm of AVs, the outcomes at stake are critical. While an out-of-order temporarily Facebook post or a lost chance at a concert ticket are unlikely to cause serious harm, an AV that fails to act appropriately is a matter of life and death.

Some simple examples help illustrate the need for potentially varying the accuracy-speed trade-off implementation in different driving environments. When detecting a pedestrian up ahead, it is not always obvious if slowing down or veering in a different direction is the right choice. On an otherwise empty country road, if an AV detects a pedestrian in the distance, it may be safer for it to take the time to more accurately determine if there is a pedestrian in its path. In a busy urban environment, it may be safer to bias toward making a decision quickly; in the case of unexpected jaywalkers, it could be catastrophic for an AV to take an extra half-second to be absolutely certain there is a pedestrian directly in front of it.[34]

Resolving the trade-off implementation for particular cases like these is out of the scope of this Article. Rather, our goal is to show that consideration of the accuracy-speed trade-off is an important contributor to the overall behavior of an AV and that determining how to appropriately implement it may need to vary by ODD.[35] We will discuss concrete policy implications of this variability in Part III. For now, the key takeaway is that, depending on environmental conditions, there will be cases in which it may be preferable for an AV to sacrifice some speed for accuracy, or some accuracy for speed. These needs will vary not just *between* AVs, but also *within* a particular AV. Rather than visualizing this requirement as a fixed point on the spectrum between accuracy and speed, as is the case for the example systems discussed earlier and in Figure 1, one should instead think of the trade-off decision as a movable dial dependent on context.[36]

---

33. *See* FAVP, *supra* note 12, at 10, 13 ("A vehicle has a separate automated vehicle system for each Operational Design Domain such that a SAE Level 2, 3 or 4 vehicle could have one or multiple systems, one for each ODD (e.g., freeway driving, self-parking, geofenced urban driving). SAE Level 5 vehicles have a single automated vehicle system that performs under all conditions.").

34. We could further complicate these examples by changing the time of day and weather conditions. The amount of light or the presence of rain could impact sensor functions. Increased traffic could require increased V2V and V2I communication.

## II. Reasoning About Similar Trade-Offs in Other High-Impact Domains

Part I set forth that the accuracy-speed trade-off is particularly complex for AVs. In this section, we discuss how the trade-off implicates broader social values that we want to balance in policymaking decisions. There is precedent for balancing similar trade-offs and their associated values in other policymaking domains. Thus, even though AVs present unique technical challenges, we can view the accuracy-speed trade-off as a regulable decision point at which policymakers can meaningfully intervene in order to promote important social values—notably, safety, efficiency, and accountability. We provide several examples from various domains, such as public health, and then show how risk assessment and management provides a particularly useful analog for thinking about how we can effectively govern the accuracy-speed trade-off for AVs. Concrete *ex ante* and *ex post* policy considerations follow from this framing, which we explore in Part III.

35. An ODD that presents a different type of "legal inaccuracy" is worth noting: In certain safety-critical situations, AVs will need to perform actions inconsistent with state traffic laws. For example, it may be necessary to drive backwards on a highway or to cross double lines in order to avoid a broken-down vehicle. NHTSA acknowledges the importance of AVs to be sufficiently flexible to implement these exceptional behaviors. *See* FAVP, *supra* note 12, at 25.

36. While we focus here on AVs, the observations we make above are more far-reaching, especially in relation to Internet of Things (IoT) systems. Such modern distributed systems appear poised to remove the remaining physical and computational barriers to ubiquitous information capture and automated, sensor-driven decision-making. As hardware sensors become cheaper to produce, processors become more powerful, and machine learning research accelerates at an explosive rate and begins to explore deployment in dynamic, mobile, distributed settings, the prospects for distributed systems also grow—making their engineering trade-offs all the more urgent to understand. For more on IoT, *see* Ken Birman et al., *Cloud-Hosted Intelligence for Real-time IoT Applications,* 53 ACM SIGOPS OPERATING SYS. REV. 1, 7–13 (2019). For more on the explosive growth of ML technology, *see generally* Jeff Dean et al., *A New Golden Age in Computer Architecture: Empowering the Machine-Learning Revolution*, 38 IEEE MICRO 21, 24 (2018).

### A. *Technical Trade-Offs Implicate Overarching Social Values*

In science and technology studies, information science, and law, it is by now axiomatic to assert that technical artifacts embed political and social values.[37] The relationships among these values are often complex. Values can be complementary but are also often in tension with each other. Moreover, low-level technical and engineering trade-offs[38] can entail system behaviors that reflect tensions among social values.[39] The trade-off between accuracy and speed is no exception; though the trade-off may seem strictly technical, depending on its specific context, it can implicate a range of normative considerations.

Revisiting our examples of the trade-off in Part I demonstrates the different values embedded in trade-off implementation choices. Social media biases toward the speed end of the trade-off spectrum. It favors a fast user experience, which is correlated with optimizing user engagement (and thus revenue).[40] Transient accuracy issues that, for example, cause temporarily out-of-order comments on a feed are a less crucial consideration for user engagement; and thus, accuracy is deprioritized in the technical implementation of the trade-off. In short, social media sites value profitability and privileging speed is the technical choice best aligned with this aim. The relative slowness of withdrawing from a bank ATM is reflective of the choice to prioritize accuracy. Accuracy in updating a balance is more important than speed. It is more important to ensure that the balance is never—even momentarily—incorrect in order to maintain customer trust.[41]

---

37. *See* Langdon Winner, *Do Artifacts Have Politics?*, 109 DAEDALUS 121, 121–122 (1980).

38. Notably, tensions between values can get mistakenly cast as all-or-nothing trade-offs, when in fact there is a spectrum of choices. *See generally* Mary Flanagan et al., *Embodying Values in Technology: Theory and Practice, in* INFORMATION TECHNOLOGY AND MORAL PHILOSOPHY 322 (J. Van den Hoven & J. Weckert, eds., 2008). For examples of complex values tensions, Stephen Holmes, *In Case of Emergency: Misunderstanding Tradeoffs in the War on Terror*, 97 CALIF. L. REV. 301, 312–18 (2009) (discussing the fallacy of conceiving of liberty and security as a binary trade-off, with privacy being treated under the umbrella term of liberty); David E. Pozen, *Privacy-Privacy Tradeoffs,* 83 CHICAGO L. REV. 221, 245–46 (2016) (concerning how different aspects of the same value can be in conflict); *See generally* DANIEL J. SOLOVE, NOTHING TO HIDE: THE FALSE TRADEOFF BETWEEN PRIVACY AND SECURITY (2011) (explaining the false trade-off between privacy and security, discussing conflicts and tensions that arise between the two and questions of how to reconcile them).

39. *See generally* BATYA FRIEDMAN & DAVID G. HENDRY, VALUE SENSITIVE DESIGN: SHAPING TECHNOLOGY WITH MORAL IMAGINATION (2019); David E. Pozen, *Privacy-Privacy Tradeoffs,* 83 CHICAGO L. REV. 221, 221–22 (2016).

40. *See* Abadi, *supra* note 24 and accompanying text.

41. Of course, if this interaction were especially slow—for example, slower than it needed to be in order to ensure accuracy—then this could decrease customer trust.

In the case of AVs, safety is the value of paramount concern. However, in addition to safety, NHTSA notes the importance of AV efficiency.[42] It is possible to significantly slow down the operation of AVs in order to improve overall safety; however, making AVs *too* slow would negate their utility from an efficiency standpoint. These overarching values of safety and efficiency clearly relate to the lower-level technical trade-off between accuracy and speed. Decisions concerning the trade-off affect an AV's behaviors—behaviors that policymakers may want to regulate if they create overarching safety risks.[43] In short, AVs cannot be safe without some notion of accuracy, and AVs cannot be efficient if they cannot make decisions quickly.[44]

Finally, in addition to the values of safety and efficiency that NHTSA specifically highlights, AVs need to be designed with accountability at the forefront. It has long been accepted in sociotechnical literature that computerization of systems can make accountability for errors elusive.[45] The tendency toward increased computerization in car technology—even apart from the advances in AV technology—is a compelling example of how accountability can be eroded in such computerized systems. As we discuss in Section III, AV design, which depends on an unprecedented degree of computerization, must take a particularly active approach toward enabling accountability. Not only will accountability-centered design facilitate after-the-fact analysis of accidents when they occur, it will also serve as a strong motivator for AV

---

However, such an implementation would likely indicate a suboptimal trade-off implementation in a centralized banking system. Concerning blockchain, biasing toward accuracy is necessary for trust. Users need to be confident that each node in the decentralized system agrees on the transaction record, reflecting each user's correct balance. For a more detailed treatment of Bitcoin and trust, *see generally* Gili Vidan & Vili Lehdonvirta, *Mine the Gap: Bitcoin and the Maintenance of Trustlessness*, 21 NEW MEDIA & SOC'Y 42, 42–59 (2019).

42. *See* FAVP, *supra* note 13, at 3 (noting that AVs have the potential to "…[U]proot personal mobility as we know it, to make it *safer* and even *more ubiquitous* than conventional automobiles and perhaps even *more efficient*…." (emphasis added)).

43. For a treatment of the subject of human values in relation to technological risk analysis in governance (and how the two cannot be cleanly separated), *see generally* SHEILA JASANOFF, THE ETHICS OF INVENTION: TECHNOLOGY AND THE HUMAN FUTURE 31–58 (2016).

44. We are not suggesting that the trade-off between normative values of safety and efficiency maps cleanly onto the trade-off between technical values of accuracy and speed. The relationship between values is more nuanced than the kinds of optimization curves used to operationalize computational concepts. However, the technical trade-off implicates, and may help us to reason formally about, the normative trade-off.

45. *See generally* Helen Nissenbaum, *Accountability in a Computerized Society*, 2 SCI. & ENG'G ETHICS 25 (1996) (explaining how a computerized technology presents barriers to accountability not present in other technological systems).

manufacturers to improve safety *ex ante*.[46] Clarifying the accuracy-speed trade-off is one way AVs can be designed to enable accountability; the trade-off serves as a design decision point with which stakeholders can engage to ensure that AV systems align with desired social values.

### *B. The Navigation of Similar High-Stakes Trade-Offs*

Lawyers, policymakers, and legislators are accustomed to reasoning about trade-offs in other high-stakes domains—including trade-offs that are similar in character to the accuracy-speed trade-off. Policymakers often need to make decisions to act (or not to act) in the face of incomplete information and must recognize that delaying action to gather more data and increase certainty can itself be a source of harm. The public health domain is rife with these sorts of decisions, as the COVID-19 pandemic has made abundantly clear.[47] Decision-making heuristics from cognitive psychology also exhibit recognition of the trade-off,[48] and countless examples can be found in the law. In fact, analogous trade-offs are so common in U.S. legal decision-making that they are an endemic feature of the legal system. U.S. civil and criminal procedure balances needs for comprehensive, conclusive fact-finding with time considerations reflected in speedy trial requirements, local filing deadlines, preliminary injunctive relief, and statutes of

46. *See id.* at 26 (". . .[H]olding people accountable for harms or risks they bring about provides strong motivation for trying to prevent or minimize them. Accountability can therefore be a powerful tool for motivating better practices, and consequently more reliable and trustworthy systems.").

47. As just one of many examples related to policymaking under uncertainty during the COVID-19 pandemic, the World Health Organization argued in 2020 that it was necessary to guarantee COVID-19 antibodies confer immunity prior to approval of antibody tests. Some medical professionals disagreed, emphasizing that swift action is important to prioritize in a pandemic, and argued that is the norm for clinicians to act on incomplete or inaccurate information in order to treat serious conditions with urgency. *See* MC Weinstein et al., *Waiting for Certainty on Covid-19 Antibody Tests—At What Cost?*, 383 NEW ENG. J. OF MED. e37, e37 (2020) ("Demanding incontrovertible evidence may be appropriate in the rarefied world of scholarly scientific inquiry. But in the context of a raging pandemic, we simply do not have the luxury of holding decisions in abeyance until all the relevant evidence can be assembled. Failing to take action is itself an action that carries profound costs and health consequences."). *See also* Merlin Chowkwanyun et al., *Beyond the Precautionary Principle: Protecting Public Health and the Environment in the Face of Uncertainty*, *in* BIOETHICAL INSIGHTS INTO VALUES AND POLICY (C.C. Macpherson, ed.) 145, 148–49 (2016) (discussing the management of the SARS outbreak in the early 2000s and balancing between protection of public health interests and loss of liberty due to quarantine).

48. *See generally* DANIEL KAHNEMAN ET AL., JUDGMENT UNDER UNCERTAINTY: HEURISTICS AND BIASES (1982) (explaining that the time-value of information is an important element in decision-making; waiting to act is itself an action, which can have more negative consequences than acting earlier on imperfect or conflicting information).

limitations.[49] These and other rules promoting judicial efficiency are, in the words of Justice Oliver Wendell Holmes, "a concession to the shortness of life." In other words, the law recognizes that there is social value both in correct resolutions and in making resolutions efficiently.[50]

Notably, regulatory agencies' approaches to risk assessment and policymaking are *themselves* representative of various implementations of the trade-off. Agencies like the FDA, EPA, and CDC are empowered to regulate risk in technically complex, high-impact domains. They reason about safety under conditions of uncertainty, often needing to balance considerations for safety with efficient decision-making and take different approaches to doing so.[51] In risk assessment, this trade-off is frequently framed in relation to *ex ante* and *ex post* interventions for mitigating risk. *Ex ante* mechanisms emphasize collecting evidence about potential risks before approving a new substance or technology, whereas *ex post* mechanisms often focus on allocating responsibility after a harm has occurred.

For example, the FDA tends to require multiple phases of clinical trials before a new drug is approved.[52] For the FDA, this *ex ante* process is deliberately slow to generate an accurate picture of a drug's safety. The agency is empowered to require drug manufacturers to submit large amounts of clinical data, such that detailed risk assessments can be carried out before new drugs become widely available.[53] In contrast, other agencies, which place greater value on efficiency, have their authority concentrated in *ex post* mechanisms. These mechanisms usually require the agencies, rather than private companies, to invest resources in acquiring safety-related data in order to determine accountability after harm has already occurred. For example, as we have described, NHTSA currently has weak *ex ante* regulatory tools for determining whether a motor vehicle is safe to drive. While the agency is able to set safety standards, it does not verify *ex ante* that manufacturers actually meet those standards. Instead, NHTSA requires manufacturers to self-certify their own cars as "safe," and its strongest authority is its ability to recall cars *ex post*, after they cross a certain threshold concerning faulty or substandard behavior.[54] The FDA and NHTSA represent just two examples, illustrating opposite choices concerning how agencies balance the values of safety and efficiency in relation to *ex ante* and *ex post* enforcement.

---

49. Richard Brooks and Warren Schwartz, *Legal Uncertainty, Economic Efficiency, and the Preliminary Injunction Doctrine,* 58 STAN. L. REV. 381, 382 (2005); *see, e.g.*, Douglas Lichtman, *Uncertainty and the Standard for Preliminary Relief*, 70 U. CHICAGO L. REV. 197, 199 (2003) (concerning reasoning about uncertainty and its relationship to deciding when to grant injunctive relief).

50. The precautionary principle and other heuristics are commonly used approaches in the law to obtain a suitable balance between efficient resolution and the best (i.e., most accurate) adjudicative outcomes. *See generally* Cass Sunstein, *Hazardous Heuristics*, 70 U. CHICAGO L. REV. 751, 752 (2003) (applying Kahneman's ideas from cognitive psychology to legal decision-making).

51. In science-based risk assessment, it is always necessary to make decisions in the face of some degree of uncertainty. To pass judgments in the face of incomplete or inaccurate information is inherent in the epistemological nature of science. *See* Karen Levy & David Merritt Johns, *When Open Data Is a Trojan Horse: The Weaponization of Transparency in Science and Governance*, 3 BIG DATA & SOC'Y 1, 4 (2016) ("Agencies charged with protecting public health and the environment must make decisions in the face of scientific uncertainty, because science by its nature is incomplete and only rarely provides precise answers to the complex questions policymakers pose."). *See also* NATIONAL RESEARCH COUNCIL, RISK ASSESSMENT IN THE FEDERAL GOVERNMENT: MANAGING THE PROCESS 11, 42–48 (1983) (explaining the relationship between uncertainty in scientific research and risk assessment and the differences in standards across agencies related to premarketing approval and post hoc mechanisms).

52. NATIONAL RESEARCH COUNCIL, *supra* note 51, at 43 (noting that premarketing approval "empower[s] an agency to require the submission of sufficient data for a comprehensive risk assessment, whereas other programs tend to leave agencies to fend for themselves in the acquisition of necessary data.").

## III. REGULATING THE TRADE-OFF WITH NEW TOOLS

The problem of regulating AVs, then, faces multiple layers of trade-offs between efficiency and accuracy. NHTSA's own approach to regulation can be understood as a balancing act between proactive *ex ante* and reactive *ex post* strategies. And the AV, as the *object* of regulation, also exhibits this trade-off in its technical implementation, as every one of the vehicle's decisions about how to behave must be made in light of the time-cost of ensuring additional certainty about the environment.

In this section, we call on the computer science research community to build tools that rigorously characterize this technical trade-off for policymakers. For the purposes of effective policymaking, these tools need to make the trade-off transparent and assessable for AVs. By doing so, NHTSA would be capable of

53. This process can take a lot of time, and is not always conducted without criticism concerning choosing "safety" over "efficiency"—notably recently concerning the approval of COVID vaccines for children. *See, e.g.*, Tara Parker-Pope, *Why Is It Taking So Long to Get a Covid Vaccine for Kids?*, N.Y. TIMES (Aug. 26, 2021), https://www.nytimes.com/2021/08/26/well/live/covid-vaccine-kids-time.html [https://perma.cc/M2V9-6Z3T].

54. ARWEN P. MOHUN, RISK: NEGOTIATING SAFETY IN AMERICAN SOCIETY 251–255 (2013) (noting that NHTSA's emphasis on *ex post* recalls came as a result of deregulation during the Reagan administration); Rabin, *supra* note 11, at 137–38 (concerning criticisms of NHTSA for its "continuous failure" to generate or adopt *ex ante* safety standards); VINSEL, *supra* note 14, at 77–101 (concerning NHTSA's emphasis on *ex post* tools); *see* FAVP, *supra* note 12, at 7–8, 48–67 (concerning the *ex ante* and *ex post* tools NHTSA has at its disposal).

*both* implementing more effective *ex ante* regulations[55] and better evaluating accountability *ex post* after accidents occur. Identifying the appropriate *ex ante* and *ex post* mechanisms for balancing safety and efficiency in AVs will require teasing out the particular low-level technical details. For concrete AV policymaking, it will be necessary to reason about what is "safe *enough*" for AV deployment, where "safe *enough*" depends in large part on the accuracy-speed trade-off.

### *A. The Trade-Off and Facilitating Democratic Governance*

In order to realize the promise of AVs, it will be necessary to appreciate the new risks and errors they will produce. Even if AVs do fulfill their goals, it is possible that the new problems they create could negate their benefits. For example, while AV analysts believe that AV systems will bring about previously unseen equity in transportation access to disenfranchised populations, this equity would be undercut—or worse, completely subverted—if the effects of novel risks disproportionately impacted those same populations. As we have argued throughout this article, understanding the accuracy-speed trade-off can help clarify *what* some of these unprecedented technical safety issues will include. We now turn our attention to *how* understanding the trade-off can concretely assist with regulating it in practice, with the dual purpose of reducing novel risks and helping ensure that unavoidable harms are not unfairly concentrated within disenfranchised groups.

To begin, we emphasize the importance of *transparency* concerning specific trade-off implementations, so that policymakers can reason effectively about resulting safety implications. If such choices are not transparent to policymakers, then manufacturers will have the responsibility to self-regulate the trade-off's safe implementation, which—as we will show throughout this section—will likely create additional risks and harms.

Mulligan and Bamberger have called attention to the danger of such policy-relevant decisions getting pushed into low-level technical implementation details. This places decision-making power in the hands of manufacturers and their engineers, which in turn evades public deliberation and has the potential to compromise broader democratic values.[56] That is, treating

55. Of course, the ability to implement these rules is contingent upon NHTSA's *ex ante* regulatory abilities being broadened.

56. *See* Deirdre K. Mulligan & Kenneth A. Bamberger, *Saving Governance-by-Design*, 106 CALIF. L. REV. 697, 738 (2018) (discussing the problems of placing decision-making power in the hands of manufacturers and providing treatment of these issues in the context of voting machines and other specific examples).

properties like the accuracy-speed trade-off as technical details—details that are irrelevant for policymakers to understand—can push important technical choices, and the broader values they implicate, out of the realm of public debate. In particular, sole control over testing and quality control processes effectively give manufacturers, not the public, the job of converting the law into concrete technical requirements, without public input or government oversight. This weakens the ability to regulate manufacturers, effectively enabling them to self-regulate consequential technical elements, which in turn can lead to the erosion of accountability.

We instead call upon the computer science research community to build tools that provide transparency about underlying trade-off implementations, which can help policymakers regulate overall AV system safety.[57] Technical transparency with respect to AVs is an urgent problem, particularly in light of concerns about the veracity of some AV manufacturers' claims and reticence to cooperate with federal regulators.[58] We believe—as do many others—that the prevalence of safety issues in existing AV technologies suggests that NHTSA should have stronger *ex ante* regulatory mechanisms within its toolkit prior to AVs' widespread deployment.[59] Tools for assessing the trade-off could assist with developing *ex ante* standards that have teeth. Moreover, even in the absence of such standards, these same tools would prove useful for analyzing accidents *ex post*, thereby strengthening NHTSA's ability to exercise recall authority.

### *B. The Trade-Off and Ex Ante Considerations*

We next discuss how tools that expose the trade-off could help clarify effective standards or pre-market approval criteria for appropriate trade-off implementations in AVs.[60] Such *ex ante* mechanisms could also have the benefit of increasing public acceptance of HAV technology.[61] While the (thus far successful) arguments against pre-market regulations for motor vehicles have contended that rules like these would unjustifiably hinder technological innovation, we contend that they would in fact encourage innovations for AV systems.

---

57. Transparency is not on its own sufficient for accountability. *See generally* Joshua A. Kroll, *Outlining Traceability: A Principle for Operationalizing Accountability in Computing Systems*, *in* FACCT '21: PROC. OF THE ACM CONF. ON FAIRNESS, ACCOUNTABILITY, AND TRANSPARENCY 758 (2021).

58. *See, e.g.*, Boudette, *supra* note 5.

59. NHTSA also acknowledges that it might need new regulatory tools (including *ex ante* tools) to effectively regulate HAVs. *See* FAVP, *supra* note 12, at 68–82; Rabin, *supra* note 9, at 138 ("[NHTSA] will have to provide both front-end and back-end oversight. It will have to set up some ex ante performance standards to guide and channel industry innovation, and it will also be crucial for NHTSA to set up effective ex post oversight (perhaps through recalls) when unanticipated risks arise from design miscalculations.").

60. Rather than a standard for one specific implementation, these standards could of course involve policies concerning how the trade-off should be handled in diverse, dynamic ODDs. These technical considerations remain an open research area. We therefore elide this complexity for clarity. *See supra* Section I.

i. Resolving Inaccuracies *Ex Ante*

A concrete example will help to illustrate how tools that characterize the trade-off might help prevent accidents: a highly publicized AV crash, in which the trade-off was not adequately considered and therefore was not implemented appropriately. In 2018, one of Uber's semi-autonomous vehicles crashed into a pedestrian in Tempe, Arizona. The crash resulted from the coincidence of several errors, but one of the most severe centrally concerned the accuracy-speed trade-off: different sensors in the AV yielded conflicting information about whether or not there was a pedestrian in front of the car.[62] The AV did not resolve that inaccuracy in time to safely apply the brakes. Instead, by the time there was agreement among the sensors that a pedestrian was present, the pedestrian had already been fatally struck.

In the National Transportation Safety Board report assessing the crash, it is clear that the AV had not implemented a robust strategy to resolve the sensor inaccuracy and reach a decision.[63] The AV remained inaccurate—that is, it failed to decide what to do—for *over six seconds*, a significant amount of time for a computer to act. During that time, the AV wrestled with the question of whether or not a pedestrian was in its path. It did not have an adequate mechanism in place to handle the uncertainty that came from inaccurate sensor data. In this case, in which there were no other cars on the road, it seems likely that slowing down to take the extra time to resolve inaccuracy—and, quite likely, to save the life of the pedestrian—would have been safe for the vehicle to do.[64]

61. *See* FAVP, *supra* note 12, at 72.

62. *See* NATIONAL TRANSPORTATION SAFETY BOARD, COLLISION BETWEEN VEHICLE CONTROLLED BY DEVELOPMENTAL AUTOMATED DRIVING SYSTEM AND PEDESTRIAN (December 2019), Report Number HWY18MH010, https://data.ntsb.gov/Docket?NTSBNumber=HWY18MH010 [hereinafter "NTSB Report"].

63. See *id*.

64. It is worth noting that this example is far more complex than the gloss we have provided in this section, in terms of both the specific sensor inaccuracies and other safety issues. Notably, the AV was highly autonomous, but not fully autonomous: there was a human back-up driver who *could,* in theory, have engaged the brakes. However, she was not paying attention. *See id.* at 3. We emphasize "could" above because, as mentioned previously, a human driver needs seventeen seconds to regain full control of This example shows that there are cases in which the implementation of inaccuracy resolution policies in AVs would lead to safer outcomes. Of course, the case of this specific Uber crash will certainly not generalize to all potential AV accidents.[65] But as a baseline, AVs need to be able to reason about the degree of inaccuracy they experience; they could detect inaccuracy issues and attempt to correct them, which the Uber AV could not do in this case. Similarly, this ability to reason about degrees of accuracy could inform the creation of concrete *ex ante* AV standards—standards requiring *how* AVs should implement policies to resolve inaccuracies when they occur in real-time.

There is some precedent for such tools in other contexts, so calling for such tools to enable *ex ante* standards-setting for AVs is not within the realm of science fiction. For example, Facebook has built tools to monitor and correct for accuracy violations. While such tools will not transfer directly to the context of AVs, they indicate a starting point for helping to design pre-market approval standards for detecting violations and determining appropriate inaccuracy resolution strategies.[66]

#### ii. Understanding the Trade-off to Promote Innovation

Rather than seeing such tools and associated standards as a hindrance to innovation—as an impediment to the wide-scale deployment of AVs—we contend that they should be considered mechanisms that promote innovation, particularly in safety-related features.[67] Manufacturers have a duty of care[68] to produce safe vehicles. In the past, this principle has encouraged a wide variety of safety innovations, including seat belts, airbags, fuel economy, headlights, and "crashworthy" glass.[69] In other words, these (once novel) features provide empirical support that accountability can

---

an AV and the human driver in this case only had six seconds to do so. *See* note 28 and accompanying text.

65. Other than the fact that the conditions of this crash reflected a very particular set of ODD parameters, Uber in general had very lax standards concerning quality control that hopefully do not reflect the practices of AV manufacturers more generally. *See* NTSB Report, *supra* note 62, at 14.

66. *See* Haifeng Yu & Amin Vahdat, *Design and Evaluation of a Continuous Consistency Model for Replicated Services*, *in* PROC. OF THE 4TH CONF. SYMP. ON OPERATING SYS. DESIGN & IMPLEMENTATION (2000) (providing a technical discussion of inconsistency measurement); Lu, et. al, *supra* note 24.

67. *See* FAVP, *supra* note 12 and accompanying text.

68. *See* Larsen v. General Motors Corp., 391 F.2d 495, 500 (8th Cir. 1968) (concerning manufacturers' duty of care for safe motor vehicle products).

69. *See* VINSEL, *supra* note 14, at 129–130; Case comment, *Torts. Products Liability. Automobile Manufacturer Has a Duty to Protect Users of Its Product against Unreasonably Dangerous Defects in Automobile Design. Grundmanis v. British Motor Corp., 308 F. Supp. 303 (E. D. Wis. 1970)*, 84 Harv. L. Rev. 1023, 1024–25 (1971).

serve as a motivator for better standards, which in turn can prompt the development of safety-related innovations.[70]

We similarly argue that, in addition to facilitating the creation of effective *ex ante* standards, developing tools for characterizing the accuracy-speed trade-off will spur innovations in AV safety that can be integrated within AV systems to prevent accidents like the Uber crash. More broadly, such tools may also prove useful for regulation in other high-impacts, real-time distributed systems domains that exhibit accuracy-speed trade-offs, such as high-frequency trading.[71]

### *C. The Trade-Off and Ex Post Considerations*

While the Uber crash shows how understanding the trade-off is useful for *ex ante* policy considerations, it also demonstrates the utility of such a tool for *ex post* crash analysis. Granular tools that can convey detailed information, such as the degree and duration of inaccuracy between sensors, will likely be necessary to adequately determine if, during the course of an accident, an AV navigated the trade-off appropriately[72] and nevertheless could not avoid a collision. Trade-off characterization tools could help untangle such issues *ex post*, and in turn would help NHTSA determine whether and when to engage their (expansive and expensive) regulatory authority for a recall—to determine if a particular anomalous behavior was an edge case within some

70. *See id.* at 299–318; *see* note 15 and accompanying text. Automakers in fact have a history of treating safety as an innovative feature. *See* Nissenbaum *supra* note 45 and accompanying text (concerning Nissenbaum accountability as a motivator for better standards); s*ee also* Mohun, *supra* note 55, at 179–83 (concerning car company marketing strategies in the 1930 and 1950s and treating safety as a marketable design feature).

71. Explaining the implications of the accuracy-speed trade-off for high-frequency trading (HFT) in detail is out of scope for this Article. In brief, HFT technology leverages the inaccuracy inherent in super-high-speed trading (which is built on distributed systems technology) for financial gain. The distributed nature of trading systems allows for inconsistencies in data between exchanges, so that trades can be executed quickly: rather than waiting for the exchanges to reconcile their information about the state of the market, trades can proceed on potentially stale or inaccurate pricing data. HFT leverages these inaccuracies through a practice called latency arbitrage (sometimes stale quote arbitrage). *See generally* Matt Prewitt, *High-Frequency Trading: Should Regulators Do More?*, 19 MICH. TELECOMM. & TECH. L. REV. 131 (2012). This area is notoriously difficult to regulate because it is hard to attribute wide-reaching problems like "flash crashes" to specific HFT trades, in part because the accuracy-speed trade-off is not rigorously categorized and employed in tools to monitor HFT trades comprehensively. For more concerning flash crashes, *see* Kristin N. Johnson, *Regulating Innovation: High Frequency Trading in Dark Pools*, 42 J. OF CORP. LAW 833, 837 (2017).

72. Importantly, the terms "appropriately" and "reasonably" are inherently subject to interpretation.

deemed error tolerance threshold, or if it is indicative of a more systemic issue that requires a recall.

i. Distinguishing Issues of Systems Inaccuracy from Other Errors

*Ex post* attribution of errors in AVs is broadly important, and is particularly relevant concerning the accuracy-speed trade-off. As discussed in Section I, the trade-off is an inherent, generally unavoidable property of distributed systems; AVs, which operate in safety-critical situations and which require decisions to be made in very short time spans, will almost always have to give up some degree of inaccuracy for efficiency, especially in cases in which safe decisions are time-critical. In cases like this that involve an accident, it will be important to be able to attribute the accident to the appropriate source: an issue of trade-off implementation or another error (or combination of errors). Without being able to distinguish these types of accident-causing source issues, manufacturers could attempt to skirt accountability by blaming bugs on the inherent accuracy-speed trade-off.

In other words, it should not be possible for manufacturers to use the existence of the trade-off as a scapegoat to obscure other errors, which could be due to negligent engineering practices.[73] It should not be possible to claim that some degree of inaccuracy is always inevitable, and then to blame this inaccuracy as an overarching, nebulous source of crash-inducing error when in fact the actual source is a problem for which manufacturers should be held accountable. This is where tools for characterizing the trade-off, in addition to helping with *ex ante* accident prevention and standards-setting, could also be helpful in *ex post* analyses of how the trade-off functioned during the time interval in which an accident occurred.

The importance of proper attribution to prevent scapegoating the trade-off becomes even more apparent in the context of the automotive industry's documented history of evading accountability. Even prior to the development of AV technology, manufacturers have leveraged the increase in computerization in car technology to conceal misbehavior. The practice is in fact so common that it has its own name: the use of so-called "defeat devices." Defeat devices use computerization to undermine NHTSA's *ex ante* authority, as computerization makes it easier to misrepresent adherence to standards during self-certification. To

73. Nissenbaum, *supra* note 45, at 34–35 (explaining the term "scape-goating the computer").

give one notorious example, in September 2015, Volkswagen was found to have used computerized devices to misrepresent accurate emissions in diesel cars, which had gone unnoticed for over a decade.[74] Without tools to rigorously characterize and expose the accuracy-speed trade-off to policymakers, the trade-off could be a similar area in which increased computerization could help manufacturers cheat standards.

### ii. Ensuring Comprehensive Monitoring and Recording

NHTSA emphasizes the importance of logging for *ex post* audits of AV anomalies: "Vehicles should record, at a minimum, all information relevant to the event and the performance of the system, so that the circumstances of the event can be reconstructed."[75] In other words, the log must record as much information as possible—ideally everything—in order to reconstitute the full state of an AV system, so that it is possible to replay the timeline during which anomalous behaviors occur.

There is a rich literature on auditing mechanisms in accountable systems, including log replay. A common concern in this literature is that it is very challenging to log all events correctly in high-speed, distributed systems.[76] This is because the accuracy-speed trade-off is necessarily implicated in distributed logging, just as it is implicated in other distributed applications. Logging, just like any other computation an AV performs, requires computing resources. While more logging could help create a more accurate picture of the state of an AV's environment, it would also necessarily consume additional computing resources—resources that would otherwise be used to actuate and control an AV's

74. Shortly after, German regulators discovered similar defeat devices in General Motors and Mercedes-Benz vehicles, which misrepresented nitrogen oxide emissions to flout European emissions caps. In 1995, it became evident that when Cadillacs activated air conditioning, a defeat device disabled emissions control, leading to misrepresenting environmental impact. *See* VINSEL, *supra* note 14, at 172–94, 292–93. Aside from such computerized defeat devices, the automotive industry has more generally demonstrated a checkered history concerning covering up safety defects during self-certification. For example, *see* AMERICAN ASSOCIATION FOR JUSTICE, *supra* note 3, at 17–21, 24, 33–34 (concerning GM concealing an ignition switch defect self-certification, which after going unnoticed for 10 years, led to the deaths of at least 124 people).

75. *See* FAVP, *supra* note 12, at 17–18.

76. In some cases, a system can take snapshots of its state and, based on that state and recorded logs, re-execute prior behavior so that auditors can re-observe it on the fly and (ideally) see where something went wrong. For more on this in relation to a deterministic state machine (notably, AVs are non-deterministic, which presents additional complexities). *See, e.g.*, Andreas Haeberlen et al., *PeerReview: Practical Accountability for Distributed Systems*, PROC. ACM SIGOPS SYMP. ON OPERATING SYS. PRINCIPLES 175–88 (2007).

movements. That is, large amounts of logging could cause inefficiencies in overall AV performance.[77]

As a result, absent further innovations in logging for high-speed, safety-critical systems like AVs,[78] AV systems will have difficulty guaranteeing that logging mechanisms capture all details related to anomalies. This puts pressure on the need to determine rigorously what NHTSA means for loggers to capture "all information relevant to the event," as recorded logs need to guarantee (at least within a degree of certainty) that they provide sufficient detail to replay incidents. Without such guarantees, logs cannot be depended on as a final, catch-all mechanism for reliable *ex post* AV auditing.

As with our examples above, logging therefore also highlights the importance of tools that transparently characterize the trade-off; logging may seem like a simple, mundane function, but in high-speed, safety-critical systems like AVs, the accuracy-speed trade-off demonstrates that the details of its implementation are a relevant policy concern. In particular, specific *ex ante* standards around how to perform logging for AVs are necessary to assure that it is possible to reliably determine accountability for errors *ex post*.

#### iii. Reducing the Cost of Expert *Ex Post* Crash Analysis

As a last example of the *ex post* policy implications of the trade-off, we tie together our previous examples of potential technical tools that facilitate *ex ante* standards-setting and *ex post* incident analysis. These tools, which will help clarify trade-off implementations to policymakers, regulators, and other stakeholders, could also be used to facilitate non-experts' ability to hold AVs accountable after accidents.

When accidents occur, given the heightened complexity and sophistication of computerized control systems in AVs, judicial or regulatory assessment of the acceptable limits of engineering capabilities in relation to alleged design defects will rely on expert assessments. As Rabin and Abraham note, "contests over

77. For example, in the consistency (i.e., accuracy) violation monitoring discussion, there is too much overhead to examine every data access command to determine staleness. *See supra* note 18 and accompanying text. One possible alternative is to probabilistically sample a subset of these commands and to test them for systems accuracy violations. *See, e.g.*, Lu et al., *supra* note 24. It is important to guarantee that this sampling procedure is sufficiently representative to reconstitute the state of system.

78. Low-latency, totally-ordered loggers remain an open research area in computing. The HFT community is particularly interested in applying them for accurate monitoring and evaluation of super-low-latency trading systems. *See, e.g.*, Cong Ding et al., S*calog: Seamless Reconfiguration and Total Order in a Scalable Shared Log*, PROC. 17TH USENIX SYMP. ON NETWORKED SYS. DESIGN & IMPLEMENTATION, 325, 325–38 (2020).

blameworthiness will be replaced by examination of esoteric alleged engineering failures that can best be regarded—both from the vantage points of administrative cost and administrative feasibility—as simply having arisen out of the operation of a motor vehicle."[79] Expertise associated with understanding these "esoteric" issues will likely require engaging multiple experts across several disciplines, which will be extremely costly, especially if it is necessary to convene these experts every time an incident occurs.

Developing tools to clarify the effects of the trade-off could prove useful when determining liability in AV-related tort actions. Moreover, by exposing appropriate details about lower-level trade-off considerations, trade-off tools could help facilitate the ability for those with less specialized expertise to reason about error tolerance and risk acceptability. In some particularly complex cases, experts will likely still be necessary for understanding low-level implementation details; however, surfacing this information at a higher level of abstraction could obviate the need for such experts in some cases, reducing the cost and burden for individuals to sue manufacturers for liability.

## CONCLUSION

The accuracy-speed trade-off is an underexplored technical concept, but one with important implications for the operation and regulation of autonomous vehicles. These technologies hold much promise and might help to ensure equity in mobility but can only do so if their risks are properly accounted for and if the engineering decisions that underlie the technology are made legible to policymakers. Doing so shifts the balance of power from manufacturers to the public by enabling effective regulation, reducing barriers to tort recovery, and ensuring that public values like safety and accountability are appropriately balanced in emerging technologies.

79. *See* Abraham and Rabin, *supra* note 9, at 143.